\documentclass{iaus}
\usepackage{graphicx}

\title[S266.~~Star cluster system early evolution] 
{Early dynamical evolution \\ of star cluster systems}

\author[Genevi\`eve Parmentier]   
{Genevi\`eve Parmentier$^{1,2}$}

\affiliation{$^1$Institute of Astrophysics \& Geophysics \\ All\'ee du
  6 Ao\^ut 17,
B-4000 Li\`ege, Belgium \\ [\affilskip]
$^2$Argelander Institut fuer Astronomie \\ Auf dem Huegel 71, D-53121
  Bonn, Germany \\ email: {\tt gparm@astro.uni-bonn.de}}

\pubyear{2010}
\volume{266}  
\pagerange{119--126}
\jname{Star Clusters - Basic Galactic Building Blocks throughout Time and Space}
\editors{R.~de Grijs \& J.~Lepine, eds.}
\begin{document}

\maketitle

\begin{abstract}
Violent relaxation -- the protocluster dynamical response to the expulsion of its residual star forming gas -- is a short albeit crucial episode in the evolution of star clusters and star cluster systems.  Because it is heavily driven by cluster formation and environmental conditions, it is a potentially highly rewarding phase in terms of probing star formation and galaxy evolution.  In this contribution I review how cluster formation and environmental conditions affect the shape of the young cluster mass function and the relation between the present star formation rate of galaxies and the mass of their young most massive cluster.  
\end{abstract}

\firstsection 
\section{Introduction}
Compact gas-embedded clusters are observed to be the formation sites of a significant fraction of stars in the local Universe.  Following star formation, feedback processes -- ranging from protostellar outflows, photoionization-driven gas overpressure, stellar winds to, eventually, supernovae -- expel the gas not yet converted into stars.  That is, the efficiency with which a cluster parent core turns its gas into stars is always smaller than 100 per cent.  Not only does gas expulsion terminate star formation, it also leaves the newly-formed cluster out-of-equilibrium, thereby forcing it into violent relaxation.  In an attempt to regain a state of equilibrium, the cluster expands, loses a fraction of its stars or, even, undergo complete disruption.  These are the {\it cluster infant weight-loss} and {\it cluster infant mortality}, respectively. 

The dynamical response of clusters to gas expulsion implies that the cluster age distribution, the cluster mass function, the ratio between the mass in clusters and the mass in stars, as well as the distribution function of cluster half-mass radii all constitute promising diagnostic tools of star cluster formation conditions.  \cite[Baumgardt \& Kroupa (2007)]{BaumgardtKroupa07} have built and made available to the community an $N$-body model grid describing the temporal evolution of cluster mass-loss and cluster spatial expansion for a wide range of local star formation efficiencies (SFE; i.e. the mass fraction of dense molecular gas that cluster-forming cores turn into stars), gas-expulsion time-scales and external tidal field impacts.  To date, this is the most comprehensive grid of model clusters through violent relaxation.  Combined to the ever increasing quality of cluster observational data-sets (including the much needed cluster half-mass radius, see e.g. \cite[Scheepmaker et al.\,2007]{Scheepmaker+07}), the time is now ripe to decipher observed properties of young star clusters so as to probe their formation and environmental conditions.

The outline is as follows.  Following a brief reminder of violent relaxation chief parameters in Section \ref{sec:parmentier:para}, I highlight how formation and environmental conditions affect the shape of the cluster mass function in Section \ref{sec:parmentier:mf}.  I also present some preliminary results about ongoing work regarding the relation between the luminosity of the brightest young cluster and the present star formation rate (SFR) of spiral galaxies, namely, the $SFR-M_V^{brightest}$ relation (Section \ref{sec:parmentier:brightest}).

\section{Gas expulsion: prime parameters}
\label{sec:parmentier:para}
In what follows, the mass of a cluster is defined as the stellar mass
enclosed within its tidal radius.  We assume that the impact of the
external tidal field arises solely from the smooth gravitational
potential of a Milky-Way-type galaxy, i.e., at this stage, we neglect
the potential impact on the early evolution of a star cluster of its
natal giant molecular cloud.  We adopt the Jacobi radius of a cluster
as an estimate of its tidal radius (see \cite[Baumgardt et al., in press]{Baumgardt+10}, their eq.~1). 

In the course of violent relaxation, the mass $m_{\rm cl}$ of a star cluster evolves rapidly and must therefore be defined as an instantaneous quantity.  It obeys:
\begin{equation} 
m_{\rm cl} = F_b(t/\tau _{\rm cross}, \varepsilon,\tau _{\rm GExp}/\tau _{\rm cross}, r_h/r_t) \times \varepsilon \times m_{\rm c}\;.
\label{eq:parmentier:mcl}
\end{equation}

In this equation, $m_c$ is the mass of the cluster-progenitor molecular core, $\varepsilon$ is the local SFE
\footnote{We explicitly assume that stars have had sufficient time to come into virial equilibrium with the gas potential (i.e. gas expulsion occurs after a few crossing-times) and hence that the local SFE matches closely the effective SFE (see \cite[Goodwin (2009)]{Goodwin09} for a discussion).}
and $F_b$ is the stellar mass fraction of the initially gas-embedded cluster still contained within the cluster tidal radius at time $t$.  When using the term '(gas-)embedded cluster mass', this refers to the stellar mass $m_{ecl} = \varepsilon \times m_{\rm c}$.  In all the work presented here the local SFE is assumed to be uniform among all cores of a given simulation (i.e. no core-mass dependent SFE).  The bound fraction $F_b$ depends on the time $t/\tau _{\rm cross}$  elapsed since gas expulsion and on the gas expulsion time-scale $\tau _{\rm GExp}/\tau _{\rm cross}$ (both expressed in unit of the molecular core crossing-time), on the local SFE $\varepsilon$ and on the impact of the external tidal field.  The latter is quantified as the ratio between the half-mass radius $r_h$ and the tidal radius $r_t$ of the gas-embedded cluster (see below for details).  The older the cluster and/or the lower the SFE and/or the quicker gas expulsion and/or the stronger the host galaxy tidal field, the lower the bound fraction $F_b$.  
Building on the motion of a supershell propagating through the cluster-forming core and collecting the residual star-forming gas, \cite[Parmentier et al.~(2008)]{PGKB08} derived an expression for the gas-expulsion time-scale $\tau _{\rm GExp}/\tau _{\rm cross}$ (their eq.~6; see also fig.~3 in \cite[Parmentier \& Fritze 2009]{ParmentierFritze09}).  Because a cluster expands following gas expulsion, the mass fraction of stars it loses depends on how compact the gas-embedded cluster is compared to its tidal radius $r_t$.  The impact of the external tidal field is therefore accounted for by the ratio $r_h/r_t$.  As will be shown in Section \ref{sec:parmentier:mf}, the cluster-forming core mass-radius relation is therefore of paramount importance to the outcome of violent relaxation since it influences both the depth of the core potential well, hence the gas-expulsion time-scale, and the sensitivity to the external tidal field of the exposed cluster (see also \cite[Parmentier 2009]{Parmentier09}).

\section{Young cluster mass functions}
\label{sec:parmentier:mf}
In this section, I show how specific combinations of the core mass-radius relation, local SFE and external tidal field lead to cluster mass functions markedly different from the core mass function and, therefore, to mass-dependent cluster infant mortality.  Cluster-forming cores are spatially resolved in the nearby regions of our Galaxy only and the slope of the core mass-radius relation remains therefore ill-determined (see \cite[Parmentier, Kroupa \& Baumgardt, subm.]{ParmKrouBaum} for a discussion).  I consider two extreme cases in turn: constant core radii and constant core surface densities.  \\

{\underline{\it Constant cluster-forming core radii and mass-varying gas expulsion time-scale}}.

Under the assumption of constant radii ($r_c =1$\,pc), a sequence of increasing core mass equates with a sequence of increasing densities and deepening potential wells.  As a result, high-mass cores undergo slower gas expulsion and retain a larger fraction of bound stars than their low-mass counterparts.  Under the assumption of a weak tidal field, \cite[Parmentier et al.~(2008)]{PGKB08} showed that the shape of the post-violent-relaxation cluster mass function is strongly sensitive to the local SFE (see their figs.~2 and 3).  While a local SFE $\varepsilon \simeq 0.4$ is conducive to a cluster mass function mirroring the core mass function (mass-independent infant mortality and weight-loss), a local SFE $\varepsilon \simeq 0.2$ results in the survival of the -- initially -- highest-mass embedded-clusters only.  For an exposed cluster to survive so low a local SFE, gas must be expelled adiabatically, a condition which is met by high-mass cores and their deep potential well.  Only those produce bound clusters at the end of violent relaxation, rendering cluster infant mortality heavily mass-dependent.  As a result, a power-law core mass function gives rise to a bell-shaped cluster mass function.  

\begin{figure}[t]
\begin{center}
\includegraphics[width=10cm]{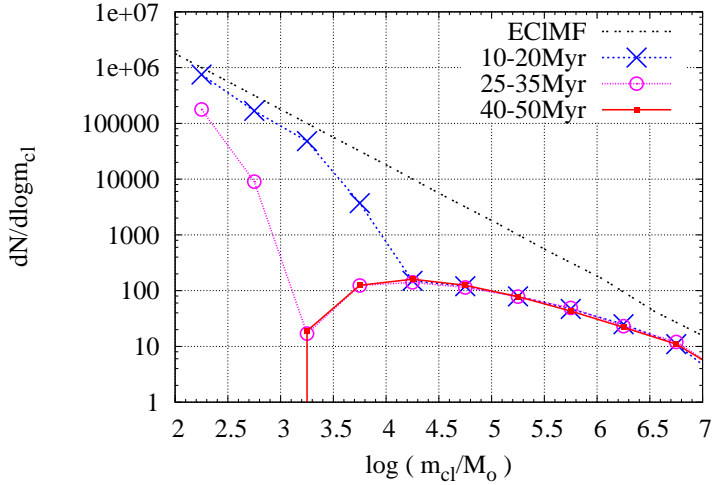} 
\caption{Time-evolution of the cluster mass function when cluster-forming core radii are constant ($r_c =1$\,pc) and the local SFE is 20 per cent.  The symbol-free dotted (black) line is the power-law embedded-cluster mass function ('EClMF').  The lines with crosses, open circles and filled squares (blue, pink and red, respectively) show the cluster mass functions integrated over the age ranges quoted in the key.}
\label{fig:parmentier:3snap}
\end{center}
\end{figure}

An aspect which \cite[Parmentier et al.~(2008)]{PGKB08} did not investigate is the time-evolution of the power-law mass function of embedded clusters in the bell-shaped cluster mass function.  The topic of evolving secularly (i.e. when violent relaxation effects have faded away) a power-law cluster mass function in a bell-shape is extensively covered in the literature.  Starting from a power-law at an age of $\sim$50-100\,Myr, internal 2-body relaxation and external tidal stripping preferentially removes low-mass clusters, thereby carving a turnover in the mass function.  As time goes by, this turnover moves towards higher cluster mass and smaller mass function amplitude.  However, the evolution of relevance here is violent relaxation instead of secular evolution and, as we shall see, the cluster mass function evolves differently.  

\begin{figure}[t]
\begin{center}
\includegraphics[width=10cm]{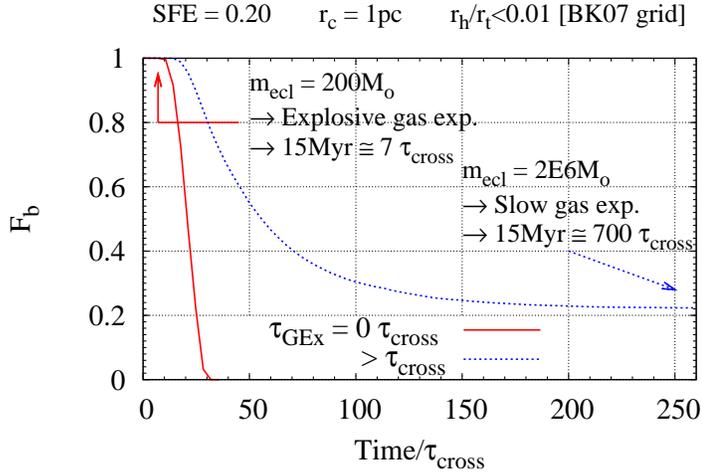} 
\caption{Time-evolution of the bound fraction of stars for an SFE of 20 per cent and two distinct regimes of cluster gas expulsion: explosive (i.e. $\tau _{\rm GExp} << \tau _{\rm cross}$) and adiabatic ($\tau _{\rm GExp} = 3\,\tau _{\rm cross}$).  Under the assumption of constant core radii, high-mass clusters have a higher density, which results in slower gas expulsion, shorter crossing-time and -- once exposed -- faster evolution than their low-mass counterparts.  }
\label{fig:parmentier:Fb}
\end{center}
\end{figure}

This is the perfect example to illustrate that the evolutionary stage of a cluster is not determined by its age in an absolute sense.  As for the gas expulsion time-scale, what matters is the age expressed in unit of a core crossing-time (see eq.~\ref{eq:parmentier:mcl}).  This implies that, keeping all other parameters alike, clusters born in cores of higher densities evolve -- once exposed -- at a rate faster than those formed out of low-density cores.  
Figure \ref{fig:parmentier:3snap} shows the embedded-cluster mass function (i.e. the core mass function shifted by $-0.7$ in $\log m$ to account for the constant 20 per cent local SFE) and snapshots of the cluster mass function at ages of $\simeq 15,\,30$ and $45$\,Myr.  The early cluster mass function (age $\simeq 15$\,Myr, dotted line with $cross$-signs) shows prominent substructures, with the cluster mass still little affected by infant weight-loss in the low- and high-mass regimes ($m_{cl} < 3,000\,{\rm M}_{\odot}$ and $m_{cl} > 2 \times 10^6\,{\rm M}_{\odot}$, respectively).  Their origin arises from how the bound fraction $F_b$ evolves with time and as a function of gas-expulsion time-scale.  In Fig.~\ref{fig:parmentier:Fb}, the solid and dotted lines show the $F_b$ time-evolutions for explosive ($\tau _{\rm GExp} << \tau _{\rm cross}$) and adiabatic (here $\tau _{\rm GExp} = 3\,\tau _{\rm cross}$) gas expulsions, respectively.  A massive embedded cluster (say, $m_{ecl} = 2 \times 10^6\,{\rm M}_{\odot}$) expels its residual gas on an adiabatic time-scale and hence retains a bound core of stars in spite of the low SFE of $20$ per cent.  Also, by virtue of the high density of its natal core, a young age of 15\,Myr equates with several hundreds of core crossing-times.  Accordingly, cluster violent relaxation is over by this young age.  This is why most of the bell-shape, which arises from clusters initially more massive than $m_{ecl} \sim 10^6\,{\rm M}_{\odot}$, is already carved by an age of 15\,Myr.

In contrast, low-mass clusters (e.g. $m_{ecl} = 200\,{\rm M}_{\odot}$) have low core densities, long core crossing-times and the same age of 15\,Myr equates with a few $\tau _{\rm cross}$ only, a stage at which the instantaneous bound fraction $F_b$ is still close to unity.  Consequently, the low-mass regime of the cluster mass function at an age of 15\,Myr sticks to the embedded-cluster mass function.  As clusters age to 50\,Myr, the cluster mass function in the high-mass regime does not evolve anymore and the low-mass regime of the mass function disappears as low-mass objects experience infant-weight loss until their eventual dissolution. \\

\begin{figure}[t]
\begin{center}
\includegraphics[width=10cm]{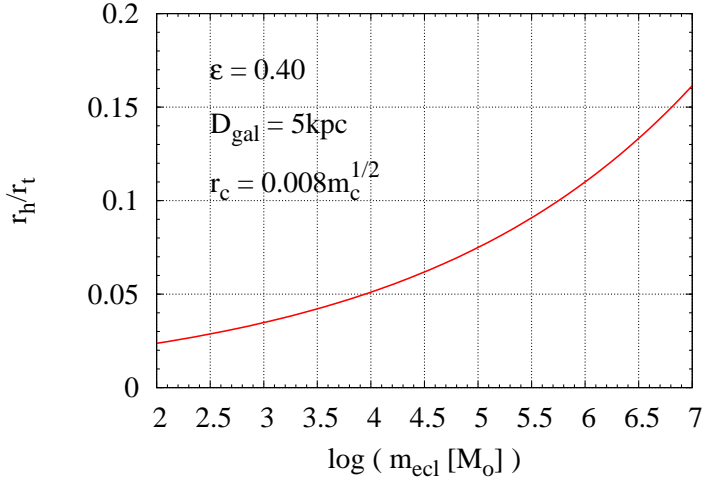} 
\caption{Dependence of the ratio between the half-mass radius $r_h$ and the tidal radius $r_t$ of the embedded cluster on its stellar mass $m_{ecl}$ for constant surface density cores.  Owing to their large spatial extent, high mass clusters show a higher sensitivity to the external tidal field and present a greater likelihood of disruption following gas expulsion.}
\label{fig:parmentier:rhrt}
\end{center}
\end{figure}

{\underline{\it Constant cluster-forming core surface densities and mass-varying tidal field impact}}.

\begin{figure}[t]
\begin{center}
\includegraphics[width=10cm]{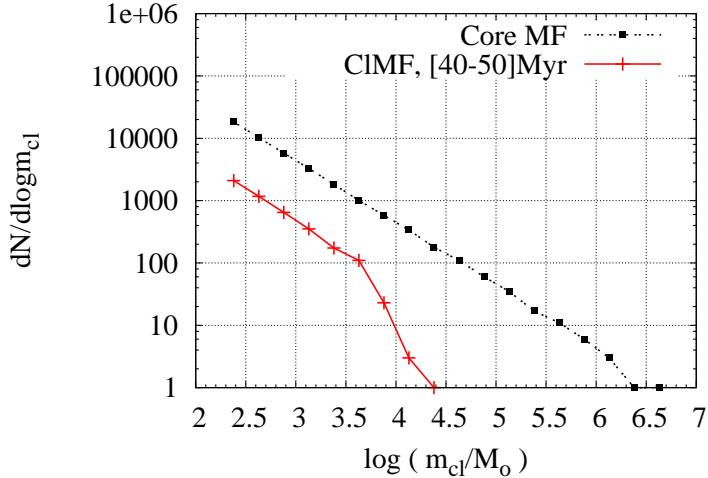} 
 \caption{Constant surface density cores: when more massive objects are more vulnerable.  The dotted (black) line is the core mass function and the solid (red) line is the cluster mass function at an age of $\simeq 50$\,Myr.  As can be understood from Fig.~\ref{fig:parmentier:rhrt}, cluster infant-mortality acts dominantly in the high-mass regime.  Model parameters are the same as in Fig.~\ref{fig:parmentier:rhrt}}
\label{fig:parmentier:mf_SB}
\end{center}
\end{figure}

Following \cite[Parmentier et al.~(2008)]{PGKB08}, molecular cores of constant surface density have an (almost) constant gas-expulsion time-scale.  Yet, the infant mortality and infant weight-loss rates of their embedded clusters are markedly mass-dependent owing to the tidal field impact.  We insist that this effect takes place even if all cores are located within the same limited region of a galaxy within which the external tidal field does not vary markedly.  The ratio between the initial half-mass radius and initial tidal radius of these embedded clusters scales as $r_h/r_t \propto m_{c}^{1/2}/m_{c}^{1/3} \propto m_{c}^{1/6}$ and, therefore, high-mass clusters have a greater sensitivity to the external tidal field since the volume of space in which they can freely expand is smaller than for low-mass clusters when compared to the initial cluster size.  This is illustrated in Fig.~\ref{fig:parmentier:rhrt} where the core mass-radius relation is $r_c = 0.008\,(m_c/1\,{\rm M}_{\odot})^{1/2}$\,pc
\footnote{
Core mass-radius relations adopted in this contribution are taken from \cite[Parmentier \& Fritze (2009)]{ParmentierFritze09}.  An update on these relations will be presented in \cite[Parmentier, Kroupa \& Baumgardt (subm.)]{ParmKrouBaum}}, 
the local SFE is $\varepsilon = 0.4$ and the galactocentric distance is $D_{gal} = 5$\,kpc in a Milky-Way-like potential.  As a result, a power-law core mass function evolves in a post-violent-relaxation cluster mass function depleted in high-mass objects (see Fig.~\ref{fig:parmentier:mf_SB}).  \\

{\underline{\it What does it all mean ?}}

The key-point to retain from this section is that the shape of the cluster mass function, the mass-radius relation of cluster-forming cores, their local SFE and the external tidal field are intimately entwined issues.  That most observed mass functions of young clusters are power-laws with a spectral index $-2$ (see \cite[Parmentier \& Gilmore (2007)]{ParmentierGilmore07} and references therein; but see also \cite[Anders et al.~(2007)]{Anders+07} for a possible turnover in the cluster mass function of the Antennae merger NGC~4038/39) therefore tells us something about cluster formation conditions which are common to different types of cluster environments.  The high sensitivity to the external tidal field of high-mass cores of constant surface density may be a hint that, contrary to theoretical expectations, the index $\beta$ of the mass-radius relation $r_c \propto m_c^{\beta}$ of cluster-forming cores is shallower than $\beta =1/2$.  More simulations are required to settle this point definitively.  In particular, greater attention must be paid to cluster radius-related quantities, e.g. the distribution function of young cluster half-mass radii and the reason why clusters lack a clear mass vs. half-mass radius correlation (\cite[Larsen 2004]{Larsen04}).  We will investigate these aspects in a forthcoming paper.     

\section{Brightest young star clusters in galaxies}
\label{sec:parmentier:brightest}
Star clusters are chronometers of major star formation events in their host galaxies (\cite[Kroupa 2002]{Kroupa02}; \cite[Kotulla, Fritze \& Anders 2008]{Kotulla+08}).  When the SFR of a galaxy increases, not only does the galaxy form a larger number of gas-embedded star clusters, the cluster mass function is also sampled up to a higher cluster mass, partly as a result of the size-of-sample effect.  Time spans in the history of galaxies characterized by vigorous star formation therefore show up prominently in cluster age histograms for two reasons: {\it (i)}~the number of clusters is larger and {\it (ii)}~these more numerous clusters also include more massive ones which experience longer dissolution time-scales and which are therefore more likely to be observed.

Observationally, this effect is best-known as a correlation between the present SFR of galaxies and the absolute visual magnitude $M_V^{brightest}$ of their brightest cluster (\cite[Larsen 2002]{Larsen02}; \cite[Weidner, Kroupa \& Larsen 2004]{WeidnerKL04}; \cite[Bastian 2008]{Bastian08}).  Building on synthetic cluster populations, \cite[Bastian (2008)]{bas08} shows that {\it if} galaxies form all their stars in gas-embedded clusters, then a universal mass fraction of 8 per cent of clusters survive the transition from their gas-embedded stage to being exposed.  

\begin{figure}[t]
\begin{center}
\includegraphics[width=10cm]{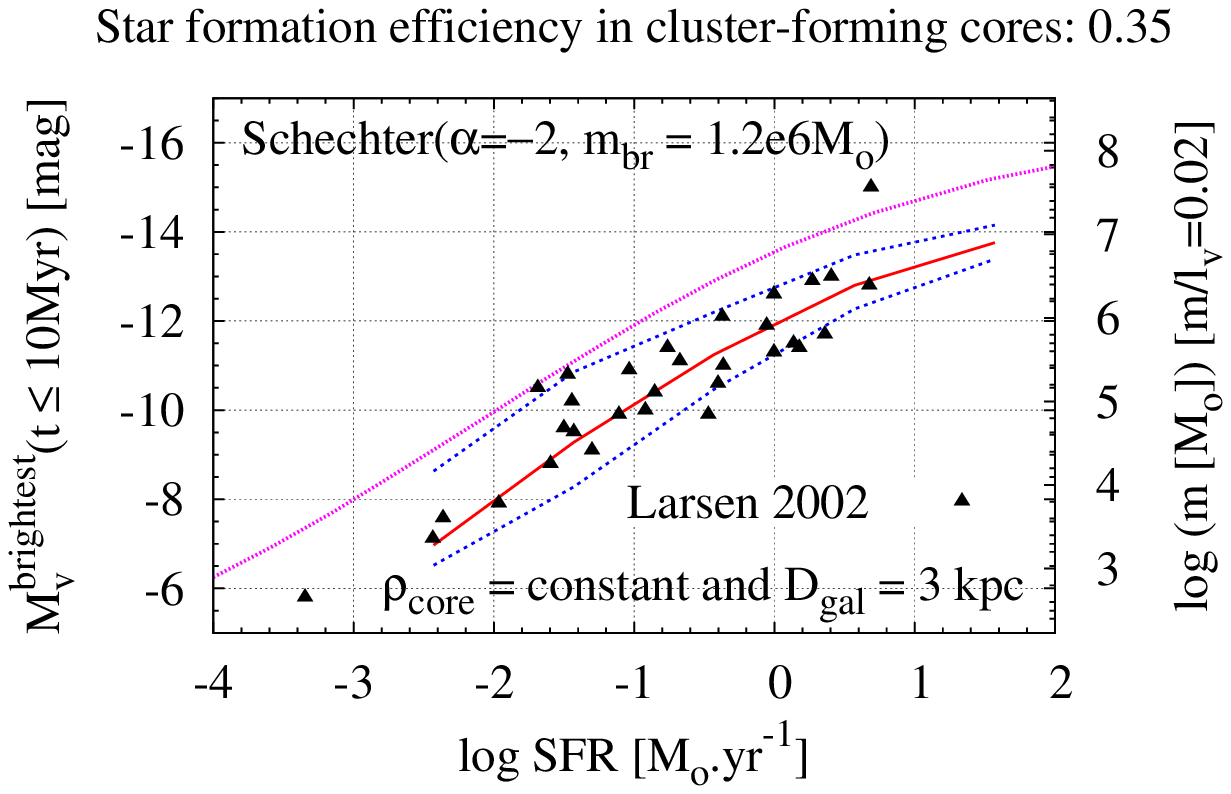} 
 \caption{Observed relation between the absolute visual magnitude of the brightest cluster
of spiral galaxies and their present SFR (filled triangles).  The dotted (pink) line is the relation between the SFR and the mass of the most massive core for the adopted Schechter core mass function ($\alpha =-2$, $m_{br}=1.2E6\,{\rm M}_{\odot}$).  Solid (red) and dashed (blue) lines depict an iso-SFE model and its $1-\sigma$ limits.  We caution that, at the exploratory stage of this model, the evolutionary time span of the model (10\,Myr) is narrower than that of the observations ($\gtrsim 100$\,Myr), thereby preventing us from drawing conclusions about the best local SFE estimate (see text for details)}
\label{fig:parmentier:SFRMv}
\end{center}
\end{figure}

That the ratio between the mass in clusters and the mass in stars depends sensitively on the mean local SFE (\cite[Parmentier \& Fritze 2009]{ParmentierFritze09}) prompted \cite[Parmentier, Kroupa \& Baumgardt (subm.)]{PKB_subm} to investigate what cluster formation conditions are the main drivers of the $SFR-M_V^{brightest}$ relation.  The absolute magnitude of a cluster depends on its age-dependent mass-to-light ratio, on the stellar mass of its parent gas-embedded cluster and on the amount of infant weight-loss this has experienced.  Section \ref{sec:parmentier:mf} demonstrates that cluster mass-related quantities are heavily shaped by cluster formation conditions.  We have therefore divided our study of how the $SFR-M_V^{brightest}$ relation responds to input parameter variations in two main parts.  Firstly, we consider the limited age range of 10\,Myr over which we explicitly assume the constancy of the integrated cluster mass-to-light ratio, that is, we investigate how the {\it mass of the most massive cluster} responds to cluster formation conditions, specifically the core mass-function, core mass-radius relation, local SFE and strength of an external tidal field (\cite[Parmentier, Kroupa \& Baumgardt, subm.]{PKB_subm}).  In a second forthcoming paper, we will extend the cluster age range to 100\,Myr and assess how mass-to-light ratio variations, hence the age of the observed brightest cluster, contribute to shape the $SFR-M_V^{brightest}$ relation.  As for the first part of the study, our findings are as follow.  The {\it vertical location} of the $SFR-M_V^{brightest}$ relation depends on the external tidal field and local SFE.  Stronger tidal fields and/or smaller SFE are conducive to smaller mass/brightness for the most massive/brightest clusters.  As for the {\it shape} of the $SFR-M_V^{brightest}$ relation, it is dictated mostly by the core mass function (see Fig.~\ref{fig:parmentier:SFRMv}) and the core mass-radius relation.  For constant surface density cores, the $SFR-M_V^{brightest}$ relation gets shallower when the SFR increases due to  the greater sensitivity to the external tidal field of high-mass embedded clusters and hence the larger amount of infant-loss they experience.  In contrast, for constant core radii, it gets steeper towards higher SFR as, in this case, higher mass cores experience slower gas expulsion and thus retain a larger bound fraction of stars. 

In the case illustrated in Fig.~\ref{fig:parmentier:SFRMv}, cluster-forming cores have a constant volume density ($r_c = 0.026\,(m_c/1\,{\rm M}_{\odot})^{1/3}$\,pc), the local SFE is 35 per cent and the synthetic cluster population is located at a galactocentric distance of 3\,kpc in a Milky-Way-like potential.  The core mass function is a Schechter function with a spectral index $-2$ and its break-mass $m_{br}$ has been adjusted so that, following star formation and infant weight-loss, the break-mass of the cluster mass function is about $\simeq 2 \times 10^5\,{\rm M}_{\odot}$ as observed for spiral galaxies (\cite[Larsen 2009]{Larsen09}).  The solid and dotted lines are the model and its $1-\sigma$ limits, respectively.  The filled triangles are the data-points from \cite[Larsen (2002)]{Larsen02} for spiral galaxies.  Figure \ref{fig:parmentier:SFRMv} suggests that an iso-SFE model provides a good fit to the ensemble of the data, provided that other conditions -- core mass-radius relation (its slope {\it and} its normalization), external tidal field -- do not vary significantly from one spiral galaxy to another.  We emphasize once again that the simulation time span (10\,Myr) is shorter than the age range of clusters shown in this plot ($\gtrsim 100$\,Myr).  Since the model vertical location is also cluster-age dependent through both the amount of infant weight-loss and the cluster mass-to-light ratio, the limited time span of our simulations prevents us from drawing firm conclusions about the best local SFE estimate.  Having now a firm handle of how cluster formation and environmental conditions affect the relation between the present SFR and the mass of the young most massive cluster, we are ready to perform simulations on a grander scale encompassing $\simeq 100$\,Myr of dynamical evolution and in which cluster mass-to-light ratio variations will be taken into account.  This will give us a {\it lower} limit on the local SFE in gas-embedded clusters of spiral galaxies, the actual value depending on the importance of the distributed mode of star formation.

\begin{acknowledgements}
I am grateful to the Scientific Organising Committee of IAU Symposium 266 for their invitation to deliver this talk and to the IAU for its financial support.  Support from the Belgian Science Policy Office in the form of a Return Grant and from the Alexander von Humboldt Foundation in the form of a Research Fellowship are acknowledged.
\end{acknowledgements}

\end{document}